\documentclass[eprint,
]{actapoly}

\usepackage[utf8]{inputenc}
\usepackage{amssymb}
\usepackage{amsmath}
\usepackage{bbm}

\hypersetup{
	colorlinks, 
	linktoc=all, 
	linkcolor=red,  
	citecolor=hyptxt,
	urlcolor=blue}
\hypersetup{
	citebordercolor=,
	filebordercolor=green,
	linkbordercolor=blue
}
\definecolor{hyptxt}{rgb}{0.7, 0.4, 0.9}

\def\calH{{\cal H }}

\def\R{\mathbb{R}}

\def\C{\mathbb{C}}

\def\ii{\mathrm{i}}

\def\lg{\langle}
\def\rg{\rangle}
\def\ud{\mathrm{d}}

\def\bu{\mathbbm{1}}
\def\bi{\widehat{\boldsymbol{\imath}}}
\def\bj{\widehat{\boldsymbol{\jmath}}}

\def\upa{| \uparrow\,\rg}
\def\dwa{| \downarrow\,\rg}

\newcommand{\ketb}[1]{|\kern.3ex#1\kern.3ex\rangle}
\newcommand{\brab}[1]{\langle\kern.3ex #1 \kern.3ex|}
\newcommand{\scalar}[2]{\langle\kern.3ex #1 \kern.3ex|\kern.3ex#2\kern.3ex\rangle}

\begin{document}

\title[Quantum angles]
{Quantum description of angles in the plane}

\author[R. Beneduci]{Roberto Beneduci}{first}
\author[E. Frion]{Emmanuel Frion}{second}
\correspondingauthor[J.-P. Gazeau]{Jean-Pierre Gazeau}{third}{gazeau@apc.in2p3.fr}

\institution{first}{Università della Calabria and Istituto Nazionale di Fisica Nucleare, Gruppo c. Cosenza, 87036 Arcavacata di Rende (Cs), Italy
}
\institution{second}{Helsinki Institute of Physics, P. O. Box 64,	FIN-00014 University of Helsinki, Finland
}
\institution{third}{Université de Paris, CNRS, Astroparticule et Cosmologie, 75013 Paris, France
}

\begin{abstract}

The real plane with its set of orientations or angles in $[0,\pi)$ is the simplest non trivial example of a  (projective) Hilbert space and provides  nice illustrations of quantum formalism. We present some of them, namely covariant integral quantization, linear polarisation of light as a quantum measurement, interpretation of  entanglement leading to the violation of Bell inequalities, and   spin one-half coherent states viewed as  two entangled angles.
\end{abstract}


\maketitle

\section{Introduction}

The formulation of quantum mechanics in a real Hilbert space has been analyzed by Stueckelberg in 1960 \cite{stueckelberg1960quantum} in order to show that the need for a complex Hilbert space is connected to the uncertainty principle. Later, Solèr \cite{soler1995characterization} showed that the lattice of elementary propositions is isomorphic to the lattice of closed subspaces of a separable Hilbert space (over the reals, the complex numbers or the quaternions). In other words, the lattice structure of propositions in quantum physics does not suggest the Hilbert space to be complex. More recently, Moretti and Oppio \cite{moretti2019quantum} gave stronger motivation  for the Hilbert space to be complex which rests on the symmetries of elementary relativistic systems. 

In this contribution, we do not address the question of the physical validity of the real Hilbert space formulation of quantum mechanics but limit ourselves to use the real 2-dimensional case, \textit{i.e.} the Euclidean plane,   as a toy model for illustrating some aspects of the quantum formalism, as  quantization, entanglement and quantum measurement. The latter is nicely  represented by the linear  polarization of light.  This real $2$-dimensional case relies on the manipulation of the two real Pauli matrices 
\begin{align}
	\sigma_1=\begin{pmatrix}
		0     &  1  \\
		1   &  0
	\end{pmatrix}\, , \quad \sigma_3=\begin{pmatrix}
		1     &  0  \\
		0   &  -1
	\end{pmatrix} \;,
\end{align}
and their tensor products, with no mention of the third, complex matrix $\sigma_2= \begin{pmatrix}
	0     &  -\ii  \\
	\ii   &  0
\end{pmatrix}$. 
As a matter of fact, many examples aimed to illustrate  tools and concepts of quantum information, quantum measurement, quantum foundations, ... (\textit{e.g.}, Peres \cite{Peres1990})  are illustrated with manipulations of these matrices. 

In \cite{bergeron2019orientations}, it was  shown that the set of pure states in the plane is represented by half of the unit circle and the set of mixed states by half the unit disk, and also that rotations in the plane rule time evolution through Majorana-like equations, all of this using only real quantities for both closed and open systems.

This paper is  a direct extension of our previous paper  \cite{beneduci2021real}, and for this reason we start the discussion by recalling some key elements of the mathematical formalism. 

\section{Background}

\subsection{Definition of POVMs}

We start with the definition of a normalized Positive-Operator Valued measure (POVM) \cite{beneduci2017joint}. It is defined as a map $F:\mathcal{B}(\Omega)\to\mathcal{L}^+_s(\mathcal{H})$ from the Borel $\sigma$-algebra of a topological space $\Omega$ to the space of linear positive self-adjoint operators on a Hilbert space $\mathcal{H}$ such that
\begin{align}
	\label{unF}
	F\left(\bigcup_{n=1}^{\infty}\Delta_n\right)=\sum_{n=1}^{\infty}F(\Delta_n)\, \quad 
	F(\Omega)=\bu \;.
\end{align}
In this definition, $\{\Delta_n\}$ is a countable family of disjoint sets in $\mathcal{B}(\Omega)$ and the series converges in the weak operator topology. If $\Omega=\mathbb{R}$, we have a real POVM. If $F(\Delta)$ is a projection operator for every $\Delta\in\mathcal{B}(\Omega)$, we recover the usual projection-valued measure (PVM).

A quantum state is defined as a non-negative, bounded self-adjoint operator with trace $1$. The space of states is a convex space and is denoted by $S(\mathcal{H})$. A quantum measurement corresponds to an affine map  $S(\mathcal{H})\mapsto\mathcal{M}_+(\Omega)$ from quantum states to probability measures, $\rho\mapsto\mu_\rho$. There is 
\cite{holevo2011probabilistic} a one-to-one correspondence between POVMs $F:\mathcal{B}(\Omega)\to\mathcal{L}_s^+(\mathcal{H})$ and affine maps $S(\mathcal{H})\mapsto\mathcal{M}_+(\Omega)$ given by $\mu_\rho(\Delta)=\mathrm{Tr}(\rho F(\Delta))$, $\Delta\in\mathcal{B}(\Omega)$.

\subsection{Integral quantization}

Quantum mechanics is usually taught in terms of projection operators and PVM, but measurements usually give a statistical distribution around a mean value, incompatible with the theory. We recall here a generalization of a quantization procedure, the \textit{integral quantization}, based on POVMs instead of PVM. The basic requirements of this programme are the following: the quantization of a classical function defined on a set $X$ must respect
\begin{enumerate}
	\item \textit{Linearity}. Quantization is a linear map $f\mapsto A_f$:
	\begin{align} \label{qmap1}
	\mathfrak{Q}:\mathcal{C}(X)\mapsto\mathcal{A}(\calH)\, , \qquad \mathfrak{Q}(f) = A_{f}\, , 
	\end{align} 
	where
	\begin{itemize}
		\item $\mathcal{C}(X)$ is a vector space of complex or real-valued functions $f(x)$
		on a set $X$, \textit{i.e.} a ``classical'' mathematical model, 
		\item $\mathcal{A}(\calH)$ is a vector space of linear operators  in some real or complex Hilbert space $\calH$, \textit{i.e.}, a ``quantum'' mathematical model, notwithstanding the question of common domains in the case of unbounded operators.  
	\end{itemize}
	\item \textit{Unity}. The map \eqref{qmap1} is such that the function $f=1$ is mapped to the identity operator $\bu$ on $\calH$.
	\item \textit{Reality}. A real function $f$ is mapped to a  self-adjoint or normal operator $A_{f}$ in $\calH$ or, at least, a symmetric operator (in the infinite-dimensional case).
	\item \textit{Covariance}. Defining the action of a symmetry group G on X by $\left(g,x\right)\in G\times X$ such as $\left(g,x\right) \mapsto g\cdot x \in X$, there is a unitary representation $U$ of $G$ such that  $A_{T(g)f}= U(g)A_f U(g^{-1})$, with $(T(g)f)(x)= f\left(g^{-1}\cdot x\right)$.
\end{enumerate}
		
Performing the integral quantization \cite{bergeron2014integral} of  a function $f(x)$ on a measure space $(X,\nu)$ boils down to the linear map:
\begin{align}
	\label{iqmap}
	f\mapsto A_{f}=\int_{X}\,{\sf M}(x)\, f(x)\,\ud\nu(x)\,,
\end{align}
where we introduce a family of operators ${\sf M}(x)$  solving the identity. More precisely, we have
\begin{align}
	\label{resunitM}
	X\ni x\mapsto\mathsf{M}(x)\,,\quad\int_{X}\,{\sf M}(x)\,\ud\nu(x)=\bu\,.
\end{align}
If the ${\sf M}(x)$ are non-negative, they provide a POVM. Indeed, the quantization of the characteristic function on the Borel set $\Delta$, $A(\chi_\Delta)$,
\begin{align}
	F(\Delta):=A(\chi_\Delta) 
	=\int_{\Delta}\,{\sf M}(x)\, \ud\nu(x) \;.
\end{align}
\noindent
is a POVM which provides a quantization procedure

$$f\mapsto A_{f}=\int_{X}\, f(x)\,dF(x).$$ 

\section{Euclidean plane as Hilbert space of quantum states}
	
\subsection{Mixed states As Density Matrices}

Density matrices act as a family of operators which can be used to perform covariant integral quantization. In the context of the Euclidean plane and its rotational symmetry, one
associates the polar angle $\phi \in [0,2\pi)$ with the unit vector $\widehat{\mathbf{u}}_{\phi}$ to define 
the pure state $|\phi\rg := \left| \widehat{\mathbf{u}}_{\phi}\right\rg$.

As shown in Figure \ref{figure quantum states}, two orthogonal pure states $\widehat{\boldsymbol{\imath}} = |0\rg$ and $\widehat{\boldsymbol{\jmath}}=\left| \dfrac{\pi}{2}  \right\rangle$ are readily identified with the unit vectors spanning the plane. In this configuration, the pure state $|\phi\rg$ is defined by an anticlockwise rotation of angle $\phi$ of the pure state $|0\rg$. Denoting the orthogonal projectors on $\widehat{\boldsymbol{\imath}}$ and $\widehat{\boldsymbol{\jmath}}$ by $| 0  \rangle \langle  0 |$ and $\left| \frac{\pi}{2} \right\rangle \left\langle  \frac{\pi}{2} \right|$ respectively, we  visualize the resolution of the identity as follows
\begin{equation}
\begin{split}
\bu = | 0  \rangle \langle  0 |  &+  \left| \frac{\pi}{2} \right\rangle \left\langle  \frac{\pi}{2} \right| \\
	 &\Updownarrow  \\
	 \left( \begin{array}{cc}
				1 & 0\\ 0 & 1 \end{array} \right) &= \left( \begin{array}{cc}
				1 & 0\\ 0 & 0 \end{array} \right)+ \left( \begin{array}{cc}
				0 & 0\\ 0 & 1 \end{array} \right) \;.
				\end{split}
	\end{equation}
	\begin{figure}[t]
	\centering
	\setlength{\unitlength}{0.1cm} 
	\begin{picture}(60,60)
		\put(10,14){\vector(1,0){30}} 
		\put(10,14){\vector(0,1){30}} 
		\put(43, 18){\makebox(0,0){$\widehat{\boldsymbol{\imath}} = |0\rangle \equiv \begin{pmatrix}
					1    \\
					0  
				\end{pmatrix}$}}
		\put(49, 44){\makebox(0,0){$|\phi\rangle = \begin{pmatrix}
					\cos\phi    \\
					\sin\phi  
				\end{pmatrix} \leftrightarrow E_{\phi}= |\phi\rg\lg \phi|$ }}  
		\put(10, 10.5){\makebox(0,0){$O$}} 
		\put(30, 2){\makebox(0,0){$\langle  0 | 0  \rangle = 1 =\left\langle \frac{\pi}{2} \right| \left.\frac{\pi}{2}\right\rangle\, , \quad \langle 0 \left| \frac{\pi}{2} \right\rangle = 0$}} 
		\put(8, 48){\makebox(0,0){$\widehat{\boldsymbol{\jmath}} = |\frac{\pi}{2}\rangle \equiv\begin{pmatrix}
					0    \\
					1  
				\end{pmatrix}$}} 
		\put(16, 32){\makebox(0,0){$1$}} 
		\put(20, 20){\makebox(0,0){$\phi$}}
		\put(14,14){\oval(10,15)[tr]}
		\thicklines 
		\put(10,14){\vector(1,2){14}} 
	\end{picture}
	\caption{The Euclidean plane and its unit vectors viewed as pure quantum states in Dirac ket notations.}
	\label{figure quantum states}
\end{figure}
Recalling that a pure state in the plane, equivalently an orientation,  can be decomposed as
	$ | \phi \rangle = \cos{\phi} \,| 0 \rangle  + \sin{\phi}\, \left| \frac{\pi}{2} \right\rangle$, with 
	$ \lg 0 | \phi\rg = \cos \phi$ and  
	$ \left\lg \frac{\pi}{2} \right|\phi\Big{\rg} =  \sin\phi$, 
it is straightforward to find the orthogonal projector corresponding to the pure state $| \phi\rangle$, 
\begin{align}
	E_{\phi} =   \begin{pmatrix}
		\cos^2\phi     & \cos\phi   \sin\phi    \\
		\cos\phi  \sin\phi    &   \sin^2\phi 
	\end{pmatrix} \;,
\end{align}
from which we can construct the density matrix corresponding to all the mixed states 
\begin{align}
	\label{specrhor}
	\rho = \left(\frac{1+r}{2}\right) E_{\phi} + \left(\frac{1-r}{2}\right) E_{\phi+\pi/2} \,, \quad 0\leq r\leq 1 \;.
\end{align}
In this expression, the parameter $r$ represents the degree of mixing. Hence the upper half-disk $(r,\phi)$, $0\leq r \leq 1$, $0\leq \phi < \pi$ is in one-to-one 
correspondence with the set of  density matrices $\rho\equiv\rho_{r,\phi}$ written as
\begin{equation}
	\label{standrhomain}
	\begin{split}
		\rho_{r,\phi}&=\frac{1}{2} \bu + \frac{r}{2}\mathcal{R}(\phi)\sigma_3\mathcal{R}(-\phi)\\&= \begin{pmatrix}
			\frac{1}{2}  + \frac{r}{2}\cos2\phi  &   \frac{r}{2}\sin2\phi  \\
			\frac{r}{2}\sin2\phi    &   \frac{1}{2}  - \frac{r}{2}\cos2\phi
		\end{pmatrix} 
	= \frac{1}{2}\left( \bu + r \sigma_{2\phi}\right)\, ,
	\end{split}
	\end{equation}
where $\mathcal{R}(\phi)=\begin{pmatrix}
	\cos\phi    &  -  \sin\phi  \\
	\sin\phi   &   \cos\phi 
\end{pmatrix}$ is a rotation matrix in the plane,  
and
\begin{align}
	\label{sigphi}
	\sigma_{\phi} &:=  \cos \phi\, \sigma_{3} +  \sin \phi\, \sigma_1 \nonumber \\ 
	&\equiv \overrightarrow{\boldsymbol{\sigma}}\cdot \widehat{\mathbf{u}}_{\phi}  = \begin{pmatrix}
		\cos \phi    &  \sin \phi  \\
		\sin \phi     & - \cos \phi
	\end{pmatrix} = \mathcal{R}(\phi)\,\sigma_3\, .
\end{align}
The observable $\sigma_{\phi}$ has eigenvalues $\{\pm 1\}$ and eigenvectors $\left| \frac{\phi}{2} \right\rangle$ and $\left| \frac{\phi +\pi}{2}   \right\rangle$ respectively.
It plays a crucial r\^ole since, as we show right after, it is at the core of both the non-commutative character and the entanglement of two quantum states of the real space. It is a typical observable used to illustrate quantum formalism \cite{Peres1990}. 

\subsection{Describing Non-commutativity and Finding Naimark Extensions through Rotations}

Let us apply integral quantization with  the real density matrices \eqref{standrhomain}. With $X= \mathbb{S}^1$, the unit circle,  equipped with the measure  $\mathrm{d}\nu(x) = \frac{\ud \phi}{\pi}$, $\phi \in [0, 2\pi)$, 
we obtain the resolution of the identity 
for an arbitrary $\phi_0$,
\begin{align}
	\label{margomegamain}
	\int_0^{2\pi} \rho_{r,\phi+\phi_0} \, \frac{\ud\phi}{\pi}= \bu\,.
\end{align}
Hence, quantizing a function (or distribution) $f(\phi)$ on the circle is done through the map
\begin{align}
	\label{qtfrhor}
		f &\mapsto A_f = \int_0^{2\pi} f(\phi) \rho_{r,\phi+\phi_0} \, \frac{\ud\phi}{\pi}
		 \nonumber \\
		 &= \begin{pmatrix}
			\lg f\rg  + \frac{r}{2}C_c\left(R_{\phi_0}f\right)  &   \frac{r}{2}C_s\left(R_{\phi_0}f\right) \\
			\frac{r}{2}C_s\left(R_{\phi_0}f\right)   &   \lg f\rg - \frac{r}{2}C_c\left(R_{\phi_0}f\right)
		\end{pmatrix} \nonumber \\
		&= \lg f\rg \,\bu + \frac{r}{2}\left[C_c\left(R_{\phi_0}f\right)  \,\sigma_3 + C_s\left(R_{\phi_0}f\right) \, \sigma_1\right]\,,
\end{align}
with $ \lg f\rg:= \frac{1}{2\pi}\int_0^{2\pi}f(\phi)\,\ud\phi$ the average of $f$ on the unit circle and $R_{\phi_0}(f)(\phi) := f(\phi-\phi_0)$. Here we have defined cosine and sine doubled angle Fourier coefficients of $f$
\begin{align}
	\label{CcCs}
	C_{\stackrel{c}{s}}(f) = \int_0^{2\pi} f(\phi) \left\lbrace\begin{array}{c}
      \cos    \\
      \sin   
\end{array}\right.2\phi \, \frac{\ud\phi}{\pi}\,.
\end{align}
In \cite{beneduci2021real}, we drew three consequences from this result. The first consequence is that, upon identification of $\R^3$ with the subspace $V_3 = \mathrm{Span}\Big{\{} e_0(\phi):=\frac{1}{\sqrt{2}},
	e_1(\phi):=\cos2\phi, 
	e_2(\phi):=\sin2\phi \Big{\}}$ in $L^2(\mathbb{S}^1,\ud\phi/\pi)$,
the integral quantization map with $\rho_{r,\phi+\phi_0}$ yields a non-commutative version of $\R^3$ :
\begin{align*}
	A_{e_0} &= \frac{\bu}{\sqrt{2}}\,,\\
	A_{e_1}&= \frac{r}{2}[\cos 2\phi_0\,\sigma_3 + \sin 2\phi_0\,\sigma_1]\equiv  \frac{r}{2} \sigma_{2\phi_0}\,, \\
	A_{e_2}&= \frac{r}{2}[-\sin 2\phi_0\,\sigma_3 + \cos 2\phi_0\,\sigma_1]\equiv  \frac{r}{2} \sigma_{2\phi_0+ \pi/2}\,.
\end{align*}
Now, the commutation rule reads
\begin{align*}
	\left[A_{e_1},A_{e_2}\right]= -\frac{r^2}{2}\tau_2\, , \quad \tau_2:= \begin{pmatrix}
		0   &  -1  \\
		1  &  0
	\end{pmatrix}=-\ii \sigma_2\,,
\end{align*}
which depends on the real version of the last Pauli matrix and on the degree of mixing.

A second consequence, typical of quantum-mechanical ensembles, is that all functions $f(\phi)$ in $V_3$ yielding density matrices through this map imply that 
\begin{align}
	\label{frho3}
	\rho_{s,\theta}= \int_0^{2\pi} \underset{f(\phi)}{\underbrace{\left[\frac{1}{2} +\frac{s}{r}\, \cos2\phi\right]}}\, \rho_{r,\phi + \theta} \, \frac{\ud\phi}{\pi} \,.
\end{align}
If $r\geq 2s$, this continuous superposition of mixed states is convex. Therefore, a mixed state is composed of an infinite number of other mixed states. This has consequences in quantum cryptography, for example, since the initial signal cannot be recovered from the output.

The third and last consequence we mention here concerns the Naimark extension of a function defined on the circle. In particular, we focus on the Toeplitz quantization of $f(\phi)$, which is a kind of integral quantization. 
In \cite{beneduci2021real}, we used this framework to show there exist orthogonal projectors from $L^2(\mathbb{S}^1,\ud\phi/\pi)$ to $\R^2$ such that for a function $f(\phi)$ the multiplication operator on  $L^2(\mathbb{S}^1,\ud\phi/\pi)$,  defined by
\begin{align}
	\label{MfV5}
	v \mapsto M_f v = fv\,,
\end{align}
maps  $M_f$ to $A_f$. They are precisely  Naimark's extensions of POVMs represented by density matrices (see \cite{beneduci2021real} for details).

\subsection{Linear Polarization of Light as a Quantum Phenomenon}

In this section, we recall that the polarization tensor of light can be expressed as a density matrix, which allows us to relate the polarization of light to quantum phenomena such as the Malus Law and the incompatibility between two sequential measurements \cite{beneduci2021real}. 

First, remember that a complex-valued  electric field for a propagating quasi-monochromatic electromagnetic wave along the $z$-axis  reads as
\begin{align}
	\overrightarrow{\mathcal{E}}(t)= \overrightarrow{\mathcal{E}_0}(t)\,e^{\ii \omega t}= \mathcal{E}_x\,\bi + \mathcal{E}_y\,\bj = \left(\mathcal{E}_{\alpha }\right)\,, 
\end{align}
in which we have used the previous notations for the unit vectors in the plane. The polarization is determined by $\overrightarrow{\mathcal{E}_0}(t)$. It slowly varies with time, and can be measured through Nicol prisms, or other devices, by measuring the intensity of the light yielded by mean values  $\propto\mathcal{E}_{\alpha}\mathcal{E}_{\beta}$, $\mathcal{E}_{\alpha}\mathcal{E}^{\ast}_{\beta}$ and conjugates. Due to  rapidly oscillating factors and a null temporal average $\lg\cdot\rg_t$, a partially polarized light is  described by the  $2\times2$ Hermitian matrix  (Stokes parameters) \cite{william1954mcmaster,schaefer2007measuring,Landau1975}
\begin{equation}
\begin{split}
&\frac{1}{J}\begin{pmatrix}
		\left\lg \mathcal{E}_{0x}\mathcal{E}^{\ast}_{0x}\right\rg_t     &  \left\lg \mathcal{E}_{0x}\mathcal{E}^{\ast}_{0y}\right\rg_t  \\
		\left\lg \mathcal{E}_{0y}\mathcal{E}^{\ast}_{0x}\right\rg_t      &  \left\lg \mathcal{E}_{0y}\mathcal{E}^{\ast}_{0y}\right\rg_t
	\end{pmatrix} \equiv \rho_{r,\phi} + \frac{A}{2}\sigma_2 \nonumber \\
 &= \frac{1+r}{2} E_{\phi} +  \frac{1-r}{2} E_{\phi + \pi/2}   + \ii \frac{A}{2}\tau_2 \;.
 \end{split}
\end{equation}
Here, $J$ describes the intensity of the wave. In the second line, it is clear that the degree of mixing $r$ 
describes linear polarization, while the parameter $A$ ($-1\leq A\leq 1$) is related to circular polarization. In real space, we have $A=0$, so we effectively describe the linear polarization of light. 
\begin{center}
	\setlength{\unitlength}{0.05cm} 
	\begin{picture}(60,60)
		\put(10,10){\line(1,0){50}}
		\put(10,10){\line(-1,0){30}}
		\put(10,10){\vector(1,0){15}}
		\put(10,10){\vector(0,1){15}}
		\put(25, 5){\makebox(0,0){$\widehat{\boldsymbol{k}}$}} 
		\put(7, 25){\makebox(0,0){$\widehat{\boldsymbol{\jmath}}$}} 
		\put(24, 18){\makebox(0,0){$\widehat{\boldsymbol{\imath}}$}} 
		\put(10,10){\line(0,1){40}} 
		\put(10,10){\line(0,-1){10}} 
		\put(10,10){\line(1,1){30}}
		\put(10,10){\line(-1,-1){10}} 
		\put(10,10){\vector(1,1){12}} 
		\put(10,10){\makebox(0,0){$\bullet$}}
		\put(10,10){\vector(-2,1){25}} 
		\put(-16,30){\makebox(0,0){$\mathrm{Re}\left(\overrightarrow{\mathcal{E}}\right)$}}
		\put(07, 50){\makebox(0,0){$y$}} 
		\put(60, 6.5){\makebox(0,0){$z$}} 
		\put(42, 37){\makebox(0,0){$x$}} 
		\thicklines 
	\end{picture}
\end{center}

We now wish to describe the interaction between a polarizer and a partially linear polarized light as a quantum measurement. We need to introduce two planes and their tensor product: the first one is the Hilbert space on which act the states $\rho^M_{s,\theta}$ of the polarizer viewed as an  orientation \textit{pointer}. 
Note that the action of the generator of rotations $\tau_2=-\ii\sigma_2$ on these states corresponds to a $\pi/2$ rotation :
\begin{align}
	\label{tau2act}
	\tau_2\rho^M_{s,\theta}\tau_2^{-1} = -\tau_2\rho^M_{s,\theta}\tau_2 = \rho^M_{s,\theta +\pi/2}\,.
\end{align}
The second plane is  the Hilbert space on which act the partially linearized polarization states $\rho^L_{r,\phi}$ of the plane wave crossing the polarizer. Its spectral decomposition  corresponds to the incoherent superposition of two completely linearly polarized waves 
\begin{align}
	\label{pointer}
	\rho^L_{r,\phi}=\frac{1+r}{2} \, E_{\phi} + \frac{1-r}{2} \, E_{\phi + \pi/2}\, . 
\end{align}
The pointer detects an orientation in the plane determined by the angle $\phi$. Through the interaction pointer-system, we generate a measurement whose time duration is the interval $I_M=(t_M-\eta, t_M +\eta)$ centred at $t_M$. The interaction is described by the  (pseudo-) Hamiltonian operator 
\begin{align}
	\label{intham}
	\widetilde{H}_{\mathrm{int}}(t)= \, g^{\eta}_M(t)\tau_2 \otimes \rho^L_{r,\phi} \,, 
\end{align}
where $g^{\eta}_M$ is a Dirac sequence with support in $I_M$, \textit{i.e.},  $$\lim_{\eta\to 0}\int_{-\infty}^{+\infty}\ud t \, f(t) \,g^{\eta}_M(t)= f(t_M)\,.$$

The interaction \eqref{intham} is the tensor product of an antisymmetric operator  for the pointer with an operator for the system which is symmetric (\textit{i.e.}, Hamiltonian).
The operator  defined  for $t_0< t_M-\eta$ as  
\begin{align}
	\label{nevop}
	U(t,t_0) &= \exp\left[\int_{t_0}^{t}\ud t^{\prime}\, g^{\eta}_M(t^{\prime})\, \tau_2 \otimes \rho^L_{r,\phi}\right] \nonumber\\
	&= \exp\left[ G_M^\eta(t)\, \tau_2 \otimes \rho^L_{r,\phi}\right]\, , 
\end{align} 
with $G_M^\eta(t)=\int_{t_0}^{t}\ud t^{\prime} \, \,g^{\eta}_M(t^{\prime})$, is a \textbf{unitary evolution operator}.  From the formula involving an orthogonal projector $P$,  
\begin{align}
	\label{expP}
	\exp(\theta \tau_2 \otimes P) = \mathcal{R}(\theta) \otimes  P + \bu\otimes (\bu-P)\, , 
\end{align}
we obtain
\begin{align}
	\label{nevop1t}
	U(t,t_0)= & \mathcal{R}\left(G_M^\eta(t)\,\frac{1+r}{2}\right) \otimes  E_{\phi} \nonumber\\
	&+ \mathcal{R}\left(G_M^\eta(t)\,\frac{1-r}{2}\right)  \otimes  E_{\phi +\pi/2}\, .
\end{align}
For $t_0<  t_M-\eta$ and $t> t_M+\eta$, we finally obtain
\begin{align}
	\label{nevop1}
	U(t,t_0)= \mathcal{R}\left(\frac{1+r}{2}\right) \otimes  E_{\phi} + \mathcal{R}\left(\frac{1-r}{2}\right)  \otimes  E_{\phi +\pi/2}\, . 
\end{align}

Preparing the polarizer in the state $\rho^M_{s_0,\theta_0}$,  we obtain the evolution  $U(t,t_0)\,\rho^M_{s_0,\theta_0} \otimes \rho^L_{r_0,\phi_0}\,U(t,t_0)^{\dag}$ of the initial state for $t> t_M+\eta$
\begin{align}
	&\nonumber  \rho^M_{s_0,\theta_0+\frac{1+r}{2}} \otimes \frac{1+r_0\cos2(\phi-\phi_0)}{2}\,E_{\phi} \\
	& \nonumber + \rho^M_{s_0,\theta_0+\frac{1-r}{2}} \otimes \frac{1-r_0\cos2(\phi-\phi_0)}{2}\,E_{\phi+\pi/2}\\
	&\nonumber + \frac{1}{4}\left(\mathcal{R}(r) +s_0\sigma_{2\theta_0 +1}\right)\otimes r_0\sin2(\phi-\phi_0)\,E_{\phi}\tau_2\\
	& - \frac{1}{4}\left(\mathcal{R}(-r) +s_0\sigma_{2\theta_0 +1}\right)\otimes r_0\sin2(\phi-\phi_0)\,\tau_2 E_{\phi}\, . 
\end{align}
Therefore, the probability for the pointer to rotate by $\frac{1+r}{2}$, corresponding to the polarization  along the orientation  $\phi$ is
\begin{align}
	\label{ pointerphi}
	&\mathrm{Tr}\left[\left(U(t,t_0)\,\rho^M_{s_0,\theta_0} \otimes \rho^L_{r_0,\phi_0}\,U(t,t_0)^{\dag}\right)\left(\bu\otimes E_\phi\right)\right] \nonumber \\
	&= \frac{1+r_0\cos2(\phi-\phi_0)}{2}\,,
\end{align} 
that for the completely linear polarization of the light, \textit{i.e.} $r_0 = 1$, becomes the familiar Malus law, $\cos^2 (\phi-\phi_0)$. Similarly, the second term gives the probability for the perpendicular orientation $\phi + \pi/2$ and the pointer rotation by $\frac{1-r}{2}$
\begin{align}
	\label{ pointerphiorth}
	&\mathrm{Tr}\left[\left(U(t,t_0)\,\rho^M_{s_0,\theta_0} \otimes \rho^L_{r_0,\phi_0}\,U(t,t_0)^{\dag}\right) \left(\bu\otimes E_{\phi+\pi/2}\right)\right] \nonumber\\
	&= \frac{1-r_0\cos2(\phi-\phi_0)}{2}\,,
\end{align}  
corresponding (in the case $r_0=1$) to the Malus law $\sin^2 (\phi-\phi_0)$.

%
%

\section{Entanglement and isomorphisms}

In this section, we develop our previous results further by giving an interpretation in terms of quantum entanglement. Previously, we described the interaction between a polarizer and a light ray as the tensor product \eqref{intham}, which is analogous to the quantum entanglement of states, since it is a logical consequence of the construction of tensor products of Hilbert spaces for describing quantum states of composite system.  In the present case, we are in presence of a remarkable sequence of vector space isomorphisms  due to the fact that $2\times 2 = 2+2$ \footnote{Remind that  $\mathrm{dim}(V\otimes W)= \mathrm{dim}V\mathrm{dim}W$ while $\mathrm{dim}(V\times W)= \mathrm{dim}V+\mathrm{dim}W$ for 2 finite-dimensional vector spaces $V$ and $W$} : 
\begin{align}
	\label{seqiso}
	\R^2 \otimes \R^2 
	\cong \R^2\times \R^2 \cong \R^2 \oplus \R^2  \cong \C^2\cong \mathbb{H}\, , 
\end{align}
where $\mathbb{H}$ is the field of quaternions. Therefore, the description of the entanglement in a real Hilbert space is equivalent to the description of a single system (\textit{e.g.}, a spin $1/2$) in the complex Hilbert space $\C^2$, or in $\mathbb{H}$. In Section \ref{qerc} we develop such an observation. 

\subsection{Bell States and Quantum Correlations}  

It is straightforward to transpose into the present setting the  1964 analysis and result presented by Bell  in his discussion about the EPR paper \cite{bell1964einstein}  and about the subsequent Bohm's  approaches  based on the assumption  of  hidden variables \cite{bohm1952suggested}. We only need to replace the Bell spin one-half particles  with the horizontal (\textit{i.e.}, $+1$) and vertical (\textit{i.e.}, $-1$) quantum orientations in the plane as the only possible issues of the observable $\sigma_{\phi}$ \eqref{sigphi},
supposing that there exists a  pointer device designed for measuring such orientations with outcomes $\pm 1$ only.

In order to define Bell states and their quantum correlations, let us  first write  the canonical, orthonormal basis of the tensor product $\R_A^2 \otimes \R_B^2$, the first factor being for system ``$A$'' and the other for system ``$B$'',  as
\begin{equation}
	\label{tensbas}
	\begin{split}
& |0\rg_A\otimes|0\rg_B\, , \quad \left|\frac{\pi}{2}\right\rg_A\otimes\left|\frac{\pi}{2}\right\rg_B\, , \\
	& |0\rg_A\otimes\left|\frac{\pi}{2}\right\rg_B\, , \quad \left|\frac{\pi}{2}\right\rg_A\otimes|0\rg_B\, .
\end{split}
\end{equation}
The states $|0\rg$ and $\left|\frac{\pi}{2}\right\rg$ pertain to $A$ or $B$, and are named ``$q$-bit'' or ``qubit'' in the standard language of  quantum information. Since they are pure states, they can be associated to a pointer measuring the horizontal (resp. vertical)  direction or polarisation described by the state $|0\rg$  (resp. $\left|\frac{\pi}{2}\right\rg$). 
		
There are four Bell pure states in $\R_A^2 \otimes \R_B^2$, namely
\begin{align}
	|\Phi^{\pm} \rg & = \frac{1}{\sqrt 2} \left( |0\rg_A\otimes|0\rg_B \pm \left|\frac{\pi}{2}\right\rg_A\otimes\left|\frac{\pi}{2}\right\rg_B\right) \, , \\
	\label{bell2}   |\Psi^{\pm} \rg&  = \frac{1}{\sqrt 2} \left( \pm |0\rg_A\otimes\left|\frac{\pi}{2}\right\rg_B + \left|\frac{\pi}{2}\right\rg_A\otimes|0\rg_B  \right)\,.
\end{align}
We say that they represent 
\emph{maximally entangled} quantum states of two qubits. Consider for instance the state $|\Phi^{+} \rg$. If the pointer associated to $A$ measures its qubit in the standard basis, the outcome would be perfectly random, with either possibility having a probability 1/2. But if the pointer associated to $B$ then measures its qubit instead, the outcome, although random for it alone,   is the same as the one $A$ gets. There is \emph{quantum correlation}. 

\subsection{Bell Inequality and Its Violation}

Let us  consider a bipartite system in the state $\Psi^{-}$. In such a state, if a measurement of the component $\sigma^A_{\phi_a}:=\overrightarrow{\boldsymbol{\sigma}}^A\cdot \widehat{\mathbf{u}}_{\phi_a}$ ($\widehat{\mathbf{u}}_{\phi_a}$ is an unit vector with polar angle $\phi_a$)  yields the value $+1$ (polarization along the direction $\phi_a/2$), then a measurement of $\sigma^B_{\phi_b}$ when $\phi_b= \phi_a$  must yield the value $-1$ (polarization along the direction $\frac{\phi_a+\pi}{2}$), and vice-versa. From a classical perspective, the explanation of such a correlation needs a predetermination by means of the existence of \textit{hidden} parameters $\lambda$ in some set $\Lambda$. Assuming the two measurements to be separated by a space-like interval, the result $\varepsilon^A\in \{-1,+1\}$ (resp. $\varepsilon^B\in \{-1,+1\}$)  of measuring $\sigma^A_{\phi_a}$ (resp. $\sigma^B_{\phi_b}$)  is then determined by $\phi_a$ and $\lambda$ only (locality assumption), not by $\phi_b$, \textit{i.e.} $\varepsilon^A= \varepsilon^A(\phi_a ,\lambda)$ (resp. $\varepsilon^B= \varepsilon^B(\phi_b ,\lambda)$). Given a probability distribution $\rho(\lambda)$ on $\Lambda$, the \underline{\textbf{classical}} expectation value of the product of the two components $\sigma^A_{\phi_a}$ and $\sigma^B_{\phi_b}$ is given by
\begin{align}
	\mathsf{P}(\phi_a,\phi_b)= \int_{\Lambda}\ud \lambda\, \rho(\lambda) \, \varepsilon^A(\phi_a ,\lambda)\,\varepsilon^B(\phi_b ,\lambda)
	\, . 
\end{align}
Since 
\begin{align}
	\int_{\Lambda}\ud \lambda\, \rho(\lambda)= 1 \, \quad \textup{and} \quad \varepsilon^{A,B}= \pm1 \;,
\end{align}
we have $-1 \leq \mathsf{P}(\phi_a,\phi_b)\leq 1$.
Equivalent predictions \underline{\textbf{within the quantum setting}} then imposes the equality between the classical and quantum expectation values:
\begin{align}
	\mathsf{P}(\phi_a,\phi_b) &= \left\lg \Psi^{-} \right| \sigma^A_{\phi_a}\otimes \sigma^B_{\phi_b}\left|  \Psi^{-} \right\rg \nonumber \\
	&= - \widehat{\mathbf{u}}_{\phi_a}\cdot \widehat{\mathbf{u}}_{\phi_b}= -\cos(\phi_a-\phi_b) \,. 
\end{align}
In the above equation,  the value $-1$ is reached at $\phi_a = \phi_b$. This is possible for $\mathsf{P}(\phi_a,\phi_a)$ only if $\varepsilon^A(\phi_a ,\lambda) = -\varepsilon^B(\phi_a ,\lambda)$. Hence,  we can write $\mathsf{P}(\phi_a,\phi_b)$ as
\begin{align}
	\mathsf{P}(\phi_a,\phi_b) &= -\int_{\Lambda}\ud \lambda\, \rho(\lambda) \, \varepsilon(\phi_a ,\lambda)\,\varepsilon(\phi_b ,\lambda)\, , \nonumber \\
	 \varepsilon(\phi,\lambda) &\equiv  \varepsilon^A(\phi,\lambda) = \pm1
	\, . 
\end{align}

Let us now introduce a third unit vector $\widehat{\mathbf{u}}_{\phi_c}$.  Due to $\varepsilon^2= 1$, we have
\begin{align}
	\mathsf{P}(\phi_a,\phi_b) - \mathsf{P}(\phi_a,\phi_c)&= \int_{\Lambda}\ud \lambda\, \rho(\lambda) \, \varepsilon(\phi_a ,\lambda)\,\varepsilon(\phi_b ,\lambda)\, \nonumber \\
	& \times \left[\varepsilon(\phi_b ,\lambda)\,\varepsilon(\phi_c ,\lambda)-1\right]\, . 
\end{align}
From this results the (baby) Bell inequality:
\begin{align*}
	&\vert\mathsf{P}(\phi_a,\phi_b) - \mathsf{P}(\phi_a,\phi_c)\vert \\ &\leq  \int_{\Lambda}\ud \lambda\, \rho(\lambda)\, \left[1-  \varepsilon(\phi_b ,\lambda)\,\varepsilon(\phi_c ,\lambda)\right] \nonumber 
	= 1+ \mathsf{P}(\phi_b,\phi_c)\, . 
\end{align*}
Hence, the validity of the existence of hidden variable(s) for justifying the quantum correlation  in the singlet state $\Psi^{-}$, and which is encapsulated by the above equation,  has the following consequence on the arbitrary triple $(\phi_a,\phi_b,\phi_c)$:
\begin{align*}
	1-\cos(\phi_b-\phi_c) \geq \left\vert \cos(\phi_b-\phi_a) - \cos(\phi_c-\phi_a)\right\vert\, . 
\end{align*} 
Equivalently, in terms of the two independent angles $\zeta$ and  $\eta$,
\begin{align*}
	\zeta= \dfrac{\phi_a-\phi_b}{2} \,, \quad
	\eta= \dfrac{\phi_b-\phi_c}{2} \nonumber \;,
\end{align*}
we have
\begin{align}
	\label{ineqsin}
	\left\vert \sin^2 \zeta - \sin^2(\eta+\zeta)\right\vert \leq \sin^2\eta\, . 
\end{align}

It is easy to find pairs $(\zeta, \eta)$ for which the inequality \eqref{ineqsin} does not hold true. For instance with $\eta=\zeta \neq 0$, \textit{i.e.}, $$\phi_b= \dfrac{\phi_a + \phi_c}{2} \;,$$
we obtain
\begin{align}
	\vert 4\sin^2\eta - 3\vert \leq 1\, , 
\end{align}
which does not hold true for all $\vert \eta\vert < \pi/4$, \textit{i.e.}, for $\left\vert \phi_a-\phi_b\right\vert = \left\vert \phi_b-\phi_c\right\vert< \pi/2$. 
Actually, we did not follow here the proof given by Bell, which is a lot more elaborate. Also, Bell considered  unit vectors in $3$-space. Restricting his proof to vectors in the plane does not make any difference, as it is actually the case in many works devoted to the foundations of quantum mechanics.

\subsection{Entanglement of two angles}\label{qerc}

Quantum entanglement is usually described by the complex two-dimensional Hilbert space $\C^2$.  As a complex vector space, $\C^2$, with canonical basis ($\mathbf{e}_1$, $\mathbf{e}_2$), has a real structure, \textit{i.e.}, is isomorphic to a real vector space which makes it isomorphic to $\R^4$, itself isomorphic to $\R^2 \otimes \R^2$. A \textbf{real} structure is obtained by considering the vector expansion
\begin{align}
	\label{}
	\C^2 \in \mathbf{v} &= z_1 \mathbf{e}_1 + z_2 \mathbf{e}_2 \nonumber \\ 
	&= x_1 \mathbf{e}_1 + y_1 \left(\ii \mathbf{e}_1\right) +  x_2 \mathbf{e}_2 + y_2 \left(\ii \mathbf{e}_2\right)\, , 
\end{align}
which is equivalent to writing $z_1= x_1 + \ii y_1$, $z_2= x_2 + \ii y_2$, and considering  the set of vectors 
\begin{align}
	\label{e1e2i}
	\left\{\mathbf{e}_1,   \mathbf{e}_2, \left(\ii \mathbf{e}_1\right), \left(\ii \mathbf{e}_2\right)\right\} 
\end{align}
as forming a  basis of $\R^4$. 
Forgetting about the subscripts $A$ and $B$ in (\ref{tensbas}),  we can map vectors in the Euclidean plane $\R^2$ to the complex ``plane''  $\C$ by
\begin{align}
	\label{R2C}
	|0\rg \mapsto 1\, , \qquad \left|\frac{\pi}{2}\right\rg \mapsto \ii\, , 
\end{align}	
which allows the correspondence between bases as
\begin{align}
	& |0\rg\otimes|0\rg = \mathbf{e}_1\, , \quad \left|\frac{\pi}{2}\right\rg\otimes\left|\frac{\pi}{2}\right\rg= - \mathbf{e}_2\, , \nonumber \\
	& |0\rg\otimes\left|\frac{\pi}{2}\right\rg =\left(\ii \mathbf{e}_1\right) \, , \quad  \left|\frac{\pi}{2}\right\rg\otimes|0\rg = \left(\ii \mathbf{e}_2\right)  \;.
\end{align}
Also, the spin of a particle in a real basis, given by the ``up'' and ``down'' states, are defined by 

\begin{align}
	\mathbf{e}_1 \equiv \upa\equiv \begin{pmatrix}
		1   \\
		0  
	\end{pmatrix}\, , \qquad    \mathbf{e}_2 \equiv \dwa\equiv \begin{pmatrix}
		0    \\
		1  
	\end{pmatrix}\, .
\end{align}

Finally, we obtain an unitary map from the Bell basis to the  basis of real structure of $\C^2$
\begin{align*}
	&\begin{pmatrix}
		|\Phi^{+} \rg    &    |\Phi^{-} \rg&  |\Psi^{+} \rg&  |\Psi^{-} \rg  
	\end{pmatrix} =\\ &\begin{pmatrix}
		\mathbf{e}_1&   \mathbf{e}_2 & \left(\ii \mathbf{e}_1\right) & \left(\ii \mathbf{e}_2\right) 
	\end{pmatrix}  
	  \frac{1}{\sqrt{2}}\begin{pmatrix}
		1    & 1 & 0 & 0  \\
		- 1   &  1 & 0& 0\\
		0 & 0 & 1 &- 1\\
		0 & 0 & 1 & 1
	\end{pmatrix}\, .
\end{align*}
In terms of respective components of vectors in their respective spaces, we have
\begin{align}
	\label{Bellmatcomp}
	\begin{pmatrix}
		x_1    \\    x_2 \\  y_1 \\  y_2 
	\end{pmatrix} = \frac{1}{\sqrt{2}}\begin{pmatrix}
		1    & 1 & 0 & 0  \\
		- 1   &  1 & 0& 0\\
		0 & 0 & 1 & -1\\
		0 & 0 & 1 & 1
	\end{pmatrix}  \begin{pmatrix}
		x^+\\  x^- \\ y^+ \\ y^-
	\end{pmatrix}\, . 
\end{align}
In complex notations, with $z^\pm = x^\pm + \ii y^\pm$, this is equivalent to 
\begin{align}
	\label{BellC2}
	\begin{pmatrix}
		z^+    \\
		z^-  
	\end{pmatrix}= \frac{1}{\sqrt{2}}\begin{pmatrix}
		1  &  -\mathsf{C}  \\
		\mathsf{C}  & 1  
	\end{pmatrix}\begin{pmatrix}
		z_1    \\
		z_2  
	\end{pmatrix}\equiv \mathcal{C}_{@} \begin{pmatrix}
		z_1    \\
		z_2  
	\end{pmatrix}\, , 
\end{align}
in which we have introduced the conjugation operator $\mathsf{C} z  = \bar z$, \textit{i.e.}, the mirror symmetry with respect to the real axis, $-\mathsf{C}$ being the  mirror symmetry with respect to the imaginary  axis.

Let us now see what is the influence of having real Bell states on Schr\"{o}dinger  cat states. The operator ``cat'' $\mathcal{C}_{@}$ can be expressed as
\begin{align}
	\label{catSU2}
	&\mathcal{C}_{@} =\frac{1}{\sqrt{2}}\left( \bu + \mathsf{F}\right)\, , \quad \mathsf{F} :=\mathsf{C} \tau_2= \begin{pmatrix}
		0   &  -\mathsf{C}   \\
		\mathsf{C}     &  0
	\end{pmatrix}\, . 
\end{align}

Therefore, with the above choice of isomorphisms, Bell entanglement in $\R^2\otimes \R^2$ is \textbf{not} represented by a simple linear superposition in $\C^2$. It involves also the two mirror symmetries $\pm \mathsf{C} $.  The operator $\mathsf{F}$ is a kind of ``flip'' whereas the ``cat'' or ``beam splitter'' operator $\mathcal{C}_{@}$  builds, using the  \textit{up} and \textit{down} basic states, the two elementary Schr\"odinger cats
\begin{align}
	\label{Scats}
	\mathsf{F}\,\upa &= \dwa\, , \quad \mathcal{C}_{@}\,\upa = \frac{1}{\sqrt{2}}(\upa + \dwa)\,, \\
	 \mathsf{F}\,\dwa &= -\upa\, , \quad \mathcal{C}_{@}\,\dwa= \frac{1}{\sqrt{2}}(-\upa + \dwa)\, . 
\end{align}

\begin{figure}[H]
\begin{center}
\includegraphics[width=2in]{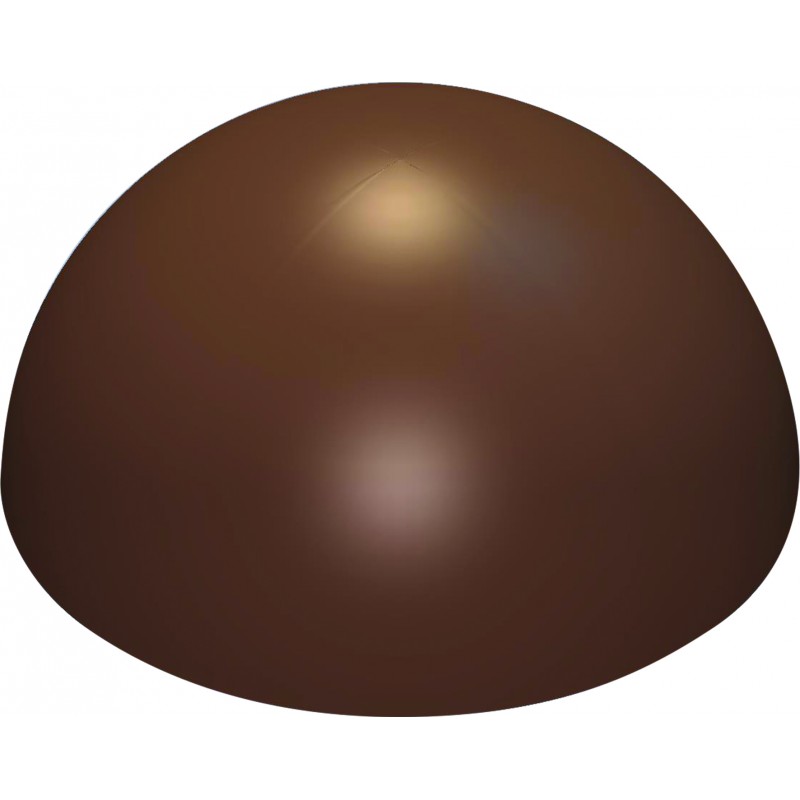}
\caption{Each point in the upper half-sphere is in one-to-one correspondence with two entangled angles in the plane. }
\label{halfsph}
\end{center}
\end{figure}

The flip operator also appears in the construction of the spin one-half  coherent states $|\theta,\phi \rg$, defined in terms of spherical coordinates $(\theta,\phi)$ as the quantum counterpart of the classical state $ \hat{\mathbf{n}}(\theta,\phi)$ in the sphere $\mathbb{S}^2$  by
\begin{equation}
	\begin{split}
&|\theta,\phi \rg=\left(\cos\frac{\theta}{2}\,\upa + e^{\ii \phi}\sin\frac{\theta}{2}\,\dwa\right) 
	\equiv \\&\begin{pmatrix}
		\cos\frac{\theta}{2}     \\
		e^{\ii \phi}\sin\frac{\theta}{2}     
	\end{pmatrix}
	=  \begin{pmatrix}
		\cos\frac{\theta}{2}   &- \sin\frac{\theta}{2} e^{-\ii \phi}  \\
		\sin\frac{\theta}{2} e^{\ii \phi}   &  \cos\frac{\theta}{2}
	\end{pmatrix}\begin{pmatrix}
		1   \\
		0 
	\end{pmatrix}\\
	&\equiv D^{\frac{1}{2}}\left(\xi^{-1}_{\hat{\mathbf{n}}}\right)\upa \,. 
	\end{split}
	 \label{operatorD} 
\end{equation}
Here, $\xi_{\hat{\mathbf{n}}}$  corresponds, through the homomorphism SO$(3)$ $\mapsto$ SU$(2)$,  to  the specific rotation $\mathcal{R}_{\hat{\mathbf{n}}}$ mapping the unit vector pointing to the north pole, $\hat{\mbox{\textbf{\textit{k}}}}=(0,0,1)$, to $\hat{\mathbf{n}}$. The operator $D^{\frac{1}{2}}\left(\xi^{-1}_{\hat{\mathbf{n}}}\right)$ represents the element  $\xi^{-1}_{\hat{\mathbf{n}}}$ of SU$(2)$ in its complex two-dimensional unitary irreducible representation. As we can see in matrix \eqref{operatorD}, the second column of $D^{\frac{1}{2}}\left(\xi^{-1}_{\hat{\mathbf{n}}}\right)$ is precisely the flip of the first one, 
\begin{align}
	\label{flipcol}
	D^{\frac{1}{2}}\left(\xi^{-1}_{\hat{\mathbf{n}}}\right) = \begin{pmatrix}
		|\theta,\phi \rg &     \mathsf{F} |\theta,\phi \rg\end{pmatrix}\,. 
\end{align}
Actually, we can learn more about the isomorphisms $\C^2 \cong \mathbb{H} \cong \R_+ \times \mathrm{SU}(2)$ through the flip and matrix representations of quaternions. In quaternionic algebra, we have the property  $\hat{\boldsymbol{\imath}}= \hat{\boldsymbol{\jmath}} \hat{\mbox{\textbf{\textit{k}}}}$ $+$ even permutations, and a quaternion $q$ is represented by  
\begin{align}
	\label{C2H}
	\mathbb{H} \ni q &= q_0 + q_1 \hat{\boldsymbol{\imath}} + q_2 \hat{\boldsymbol{\jmath}} + q_3 \hat{\mbox{\textbf{\textit{k}}}}  \nonumber \\
	&= 
	q_0  + q_3 \hat{\mbox{\textbf{\textit{k}}}}  + \hat{\boldsymbol{\jmath}} \left(q_1 \hat{\mbox{\textbf{\textit{k}}}}  + q_2 \right) \nonumber \\
	& \equiv \begin{pmatrix}
		q_0  + \ii q_3       \\
		q_2 + \ii q_1   
	\end{pmatrix} \equiv Z_q \in \C^2\, ,
\end{align}
after identifying  $ \hat{\mbox{\textbf{\textit{k}}}}\equiv \ii$ as both are roots of $-1$.  
Then the flip appears naturally in the final identification $\mathbb{H}\cong \R_+ \times \mathrm{SU}(2)$ as
\begin{equation}
	\label{HRSU2 }
	q\equiv \begin{pmatrix}
		q_0  + \ii q_3     & - q_2 + \ii q_1   \\
		q_2 + \ii q_1     &  q_0  - \ii q_3
	\end{pmatrix} = \begin{pmatrix}
		Z_q     &  \mathsf{F} Z_q  
	\end{pmatrix}\,. 
\end{equation}
Let us close this article with a final remark on spin-$1/2$ coherent states as vectors in $\R_A^2 \otimes \R_B^2$.  The ``cat states''  in $\C^2$   given by \eqref{operatorD} and equivalently viewed as $4$-vectors in $\mathbb{H}\sim \R^4$ as  
\begin{align}
	\label{entangstates}
	|\theta,\phi \rg  \mapsto  \begin{pmatrix}
		\cos\frac{\theta}{2}     \\
		-  \sin\frac{\theta}{2} \cos \phi\\
		\sin\frac{\theta}{2} \sin\phi\\
		0 
	\end{pmatrix}\, , 
\end{align} 
are represented as entangled states in $\R_A^2 \otimes \R_B^2$ by
\begin{equation}
\begin{split}
|\theta,\phi \rg = &\cos\frac{\theta}{2}  |0\rg_A\otimes|0\rg_B\ -  \sin\frac{\theta}{2} \cos \phi\left|\frac{\pi}{2}\right\rg_A\otimes\left|\frac{\pi}{2}\right\rg_B  \nonumber \\
	&+ \sin\frac{\theta}{2} \sin\phi|0\rg_A\otimes\left|\frac{\pi}{2}\right\rg_B + 0 \left|\frac{\pi}{2}\right\rg_A\otimes|0\rg_B \;.
\end{split}
	\end{equation}
Therefore, we can say that two entangled angles in the plane can be viewed as a point in the upper half-sphere $\mathbb{S}^2/\mathbb{Z}_2$
 in $\R^3$ shown in Figure \ref{halfsph}. 
 
\section{Conclusions}

Integral quantization is a quantization scheme constructed on Positive Operator-Value Measures. When applied to a two-dimensional real space, it allows for a description of quantum states as pointers in the real unit half-plane. We recalled in this paper that in this case, a family of density matrices is sufficient to perform this kind of quantization  as it describes all the mixed states in this space. Furthermore, a density matrix in a two-dimensional real space depends on the usual observable $\sigma_{\phi} = \begin{pmatrix}
	\cos \phi    &  \sin \phi  \\
	\sin \phi     & - \cos \phi
\end{pmatrix}$, which captures the essence of non-commutativity in real space. As a consequence, commutation relations are expressed in terms of the real matrix $\tau_2$, which serves as the basis to the description of quantum measurement. 

We provide an illustration considering linearly-polarized light passing through a polarizer. The pointer, associated with $\tau_2$, can rotate by an angle $(1\pm r)/2$ with $r$ the degree of mixing of the density matrix, with a probability given by the usual Malus' laws \eqref{ pointerphi} and \eqref{ pointerphiorth}. We extended the analysis by showing that the interaction between a polarizer and a light ray is equivalent to the quantum entanglement of two Hilbert spaces. Orientations in the plane have only two outcomes ($\pm1$), which are the possible issues of $\sigma_{\phi}$. We showed that for a general bipartite system, the classical and quantum measurement of $\sigma_{\phi}$ deny the existence of local hidden variables, resulting in the well-known violation of Bell inequalities, here given by \eqref{ineqsin}. Finally, we demonstrated that the isomorphism $\mathbb{C}^2 \simeq \mathbb{R}^4$ allows to write Bell states in real space, with the introduction of the ``flip'' operator \eqref{catSU2}. This operator is necessary for constructing spin one-half coherent states, that we can fully describe by a set of orientations in $\mathbb{R}^3$, as shown in \eqref{entangstates}.

\begin{acknowledgements}
R.B. The present work was performed under the auspices of the GNFM (Gruppo Nazionale di Fisica Matematica).
EF thanks the Helsinki Institute of Physics (HIP) for their hospitality.
\end{acknowledgements}

\bibliographystyle{actapoly}
\bibliography{biblio}

\clearpage
\appendix
\onecolumn

\end{document}